\documentclass[journal]{IEEEtran}
\usepackage{ifpdf}

\ifCLASSINFOpdf
  \usepackage[pdftex]{graphicx}
  \graphicspath{{../figures/}}
\else
\fi
\usepackage{amsmath}
\usepackage{algorithmic}

\usepackage{array}
\usepackage{fixltx2e}
\usepackage{url}

\usepackage[T1]{fontenc}
\usepackage{bm}
\usepackage[cmintegrals]{newtxmath}
\usepackage{mathtools, cuted}
\usepackage{color,soul}

\begin{document}
%
\title{Joint Superchannel Digital Signal Processing for Ultimate Bandwidth Utilization}
%
%
%

\author{Mikael~Mazur~\IEEEmembership{Student~Member~OSA,~IEEE}, ~Jochen~Schr{\"o}der~\IEEEmembership{Senior~Member~OSA,~Member~IEEE},
~Magnus~Karlsson~\IEEEmembership{Fellow~OSA,~Senior~Member~IEEE} and ~Peter~A.~Andrekson,~\IEEEmembership{Fellow~OSA,~IEEE}
\thanks{M. Mazur, J. Schr{\"o}der, M. Karlsson and P. A. Andrekson are with the Photonics Laboratory, Fibre Optic Communication Research Centre (FORCE), Department of Microtechnology and Nanoscience, Chalmers University  of Technology, Gothenburg  SE-412 96, Sweden (e-mail: mikael.mazur@chalmers.se; jochen.schroeder@chalmers.se; peter.andrekson@chalmers.se; magnus.karlsson@chalmers.se). 
\newline Copyright (c) 2019 IEEE. Personal use of this material is permitted.  However, permission to use this material for any other purposes must be obtained from the IEEE by sending a request to pubs-permissions@ieee.org.}}


%
%

\markboth{Journal of \LaTeX\ Class Files,~Vol.~XX, No.~XX, October~2019}%
{Shell \MakeLowercase{\textit{et al.}}: Bare Demo of IEEEtran.cls for IEEE Journals}

\markboth{}{}

%



\maketitle
\begin{abstract}
Modern optical communication systems transmit multiple frequency channels, each operating very close to its theoretical limit. 
The total bandwidth can reach 10\,THz limited by the optical amplifiers. 
Maximizing spectral efficiency, the throughput per bandwidth is thus crucial.
Replacing independent lasers with an optical frequency comb can enable very dense packing by overcoming relative drifts. 
However, to date, interference from non-ideal spectral shaping prevents exploiting the full potential of frequency combs. 
Here, we demonstrate comb-enabled multi-channel digital signal processing, which overcomes these limitations. 
Each channel is detected using an independent coherent receiver and processed at two samples-per-symbol. 
By accounting for the unique comb stability and exploiting aliasing in the design of the dynamic equalizer, we show that the optimal spectral shape changes, resulting in a higher signal to noise ratio that pushes the optimal symbol rate towards and even \emph{above} the channel spacing, resulting in the first example of frequency-domain super-Nyquist transmission with multi-channel detection for optical systems. 
The scheme is verified both in back-to-back configuration and in single span transmission of a 21 channel superchannel originating from a 25\,GHz-spaced frequency comb. 
By jointly processing 3 wavelength channels at a time, we achieve spectral efficiency beyond what is possible with independent channels. At the same time, one significantly relaxes the hardware requirements on digital-to-analog resolution and bandwidth, and well as filter tap numbers. 
Our results show that comb-enabled multi-channel processing can overcome the limitations of classical dense wavelength division multiplexing systems by enabling tighter spacing to reach the ultimate spectral efficiency in optical communications. 
\end{abstract}


%
\IEEEpeerreviewmaketitle

\section{Introduction}
Long-haul optical fiber communication systems are now connecting the world. 
Wavelength division multiplexing (WDM), the ability to transmit many channels on different wavelengths simultaneously has enabled
enormous parallelization to increase throughput rates which now often exceed 10s of Tb/s with >100 channels~\cite{Winzer2018}. However, the overall bandwidth
of fiber systems is limited by the gain of the erbium-doped fiber amplifiers (EDFAs) that periodically amplify the channels
along a link. Optimizing spectral efficiency (SE), i.e. the number of bits per unit time and frequency is therefore key to further 
increase fiber throughputs. 

The approach followed over the last decade has been to increase the per-channel SE, via advanced modulation formats. These necessitate 
the digital signal processing (DSP) with multiple-input multiple-output (MIMO) equalization~\cite{Faruk2017} and strong forward error correction (FEC)~\cite{Leven2014} to ensure successful data transmission.
However, individual channels are now so close to the theoretical limits that a further improvement of the per-channel SE is extremely
difficult for a given signal-to-noise ratio (SNR) and therefore increasing data-rates implies a throughput-reach
trade-off~\cite{Bosco2019}.

On the other hand, gaps between individual wavelength channels, so-called guard-bands, waste significant resources and thus system throughput.
Guard-bands are needed to cope with laser frequency drifts and, importantly, to enable optical routing in networks, using
so-called reconfigurable optical add-drop multiplexers (ROADMs)~\cite{Morea2014}. Much research has been spent on more holistic 
networking approaches that minimize inter-channel guard-bands, which lead laying out wavelength channels on flexible grids, instead of 
traditional fixed frequency spacings~\cite{Rafique2013}. A central component of these more flexible networks are so-called \emph{superchannels}, multi-Tb/s 
units composed of several tightly-spaced wavelength channels~\cite{Winzer2017}. 

To maximize SE, spectral gaps between channels need to be minimized. In superchannels based on individual lasers, guard-bands 
of a few GHz are needed to avoid catastrophic overlap due to frequency drifts~\cite{Liu2018,Rahn2018}. These gaps can be reduced using frequency locked 
sources such as optical frequency combs, allowing for sub-GHz guard-bands~\cite{Millar2016,mazur2018enabling}. However, even such small gaps correspond to a significant SE-loss,
which is the largest SE-loss for superchannels. 
A further reduction of spectral gaps, using a single-channel approach
is prevented by the finite pulse-shaping roll-off-factor -- which sets the tails of the spectrum-- and transceiver impairments. The limit of so-called \emph{Nyquist-WDM} \cite{Bosco2011a}, at zero roll-off, with perfectly rectangular channel spectra, is therefore only possible in theory.

However, by using carefully selected pulse shapes, it was shown by Mazo \cite{Mazo1975} that transmission with symbol rates \emph{above} the spectral width, so called \emph{super-Nyquist} transmission, is indeed possible, provided sufficient interference cancellation is performed. Several examples of that in optical links have been demonstrated \cite{Yu2013,Igarashi2014,Secondini2015}. These had in common that the intersymbol interference has been cancelled in the time domain, by processing channels \emph{independently}.
By instead processing several wavelength channels \emph{jointly}, as recently theoretically shown \cite{Jana2018}, a number of practical benefits can be accomplished; In addition to zero (or in the super-Nyquist case negative) guard-bands, the roll-off factors can be reasonably large which avoids excessive number of filter taps. Additionally, hardware requirements on bandwidth and resolution of DACs can be significantly reduced.

In this paper we present the first transmission experiment with frequency-domain interference cancellation, enabling super-Nyquist spectral efficiencies with a simultaneous significant reduction of complexity. We introduce a multi-channel DSP algorithm that enables the elimination of inter-channel guard-bands to push superchannel SE to the ultimate limits. In contrast to \cite{Jana2018}
our technique extends the dynamic MIMO equalizer used for polarization demultiplexing to remove inter-channel cross-talk by capitalizing on the frequency-locked nature of optical frequency combs.
Importantly, it exploits aliasing to work with two-samples per symbol, without requiring further upsampling or an additional equalization stage. 
We investigate the performance of our technique with a 21 channel superchannel originating from a 25\, GHz optical frequency comb in back-to-back (B2B) measurements for varying symbol-rates and roll-off factors. 
The approach is further verified by transmission experiments in 80\,km standard single-mode fiber (SMF). 
Our results show that the multi-channel DSP shifts the optimal guard-band towards zero by changing the optimal spectral pulse shape to improve signal-to-noise ratio and thereby pushing system throughput to the optimum.

\section{Dense Comb-based Superchannels}
\label{sec:joint_dsp}
Considering a superchannel with  $2N+1$ channels of equal intensity $E_s$, the received baseband signal for channel $l$ can be written according to
\begin{equation}
\label{eq:superchannel}
    r_l(t) = E_s \sum_{i=-N}^{N} \sum_{k=-\infty}^{\infty}s_{i,k} g_{i,l}(t-kT)\exp(j(\Omega_i t +\phi_i)) + n(t),
\end{equation}
with $s_{i,k}$ denoting the kth transmitted symbol on the ith channel and $n(t)$ being modelled as additative white Gaussian noise (AWGN) from both amplifier noise and noise from photo detection. In Eq.~\ref{eq:superchannel}, $\phi_i$ denotes the phase and $\Omega_i$ the frequency offset for carrier $i$. The presence of $\Omega_i$ generalizes Eq.~\ref{eq:superchannel} compared to the previously presented multi-channel DSP in~\cite{Pan:12}, as this assumed perfect frequency synchronization, corresponding to using the same comb in both the transmitter and the receiver. 
The signal envelope $g_{i,l}(t)$ denotes the response from channel $i$ after being demultiplexed with the receiver demultiplexer for channel $l$. It also contains the combined filter response of the transmitter pulse shaping filter (assumed to be a root-raised cosine filter (RRC)) and unknown responses such as component bandwidth limitations. 
As we in this work only focus on linear DSP, we neglect any non-linear distortions in Eq.~\ref{eq:superchannel}. 

In a conventional system, sufficient guard-band between each channel is typically assumed. In this case, there is no cross-talk between adjacent channels in Eq.~\ref{eq:superchannel}, or equivalently
\begin{equation}
    \int_{-\infty}^{\infty}g_{i,l}(t-kT) g^*_{i,l}(t-kT)dt = 0 \quad \text{if} \quad i\neq l,
\end{equation}
and each channel can be processed independently without any performance degradation. However, intersymbol interference (ISI) can still be present and a dynamic equalizer is typically used to minimize the penalty from ISI while avoiding noise enhancement. In frequency domain, this can be seen as weighting each spectral band to recover the maximum signal energy. 
The ISI require the equalizer to have a given temporal memory (length of the filtering tap vectors) but the spectral width required is only dictated by the shape of the channel under consideration.

However, this assumption is not valid for dense comb-based superchannels as significant cross-talk from neighboring channels is present, causing interchannel interference (ICI). 
Assuming that the cross-talk is limited to the closest neighbors, Eq.~\ref{eq:superchannel} can be re-written according to
\begin{equation}
\label{eq:Superchannel11}
\begin{split}
    r_l(t) = &E_s  \sum_{k=-\infty}^{\infty}s_{l,k} g_{l,l}(t-kT)\exp(j(\Omega_lt + \phi_l)) + n(t) + \\ E_s  &\sum_{k=-\infty}^{\infty}s_{l\pm 1,k} g_{l\pm 1,l}(t-kT)\exp(j(\Omega_{l\pm 1} t + \phi_{l\pm 1} )).
    \end{split}
\end{equation}
Moreover, while ICI is traditionally modelled as AWGN included in the noise term $n(t)$, Eq.~\ref{eq:Superchannel11} shows that this cross-talk is dependent on the transmitted symbols and spectral shape of the side channels. It is therefore not accurately modelled by AWGN.

\section{Comb-Enabled Joint Multi-channel DSP}
\begin{figure}[t]
\centering\includegraphics[width=0.8\linewidth]{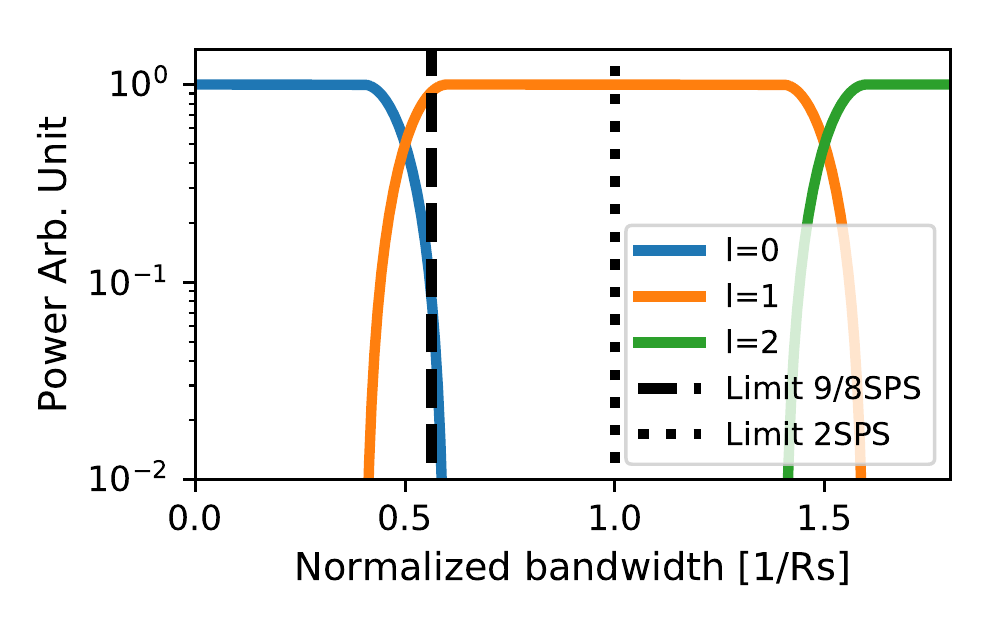}
\caption{\label{Fig:Oversamplig}Illustration of the effect from limited oversampling on equalization bandwidth. The illustrated system use raised cosine pulse shaping with $\beta=0.2$ and channels spaced at the Nyquist rate. }
\end{figure}

To improve performance with respect to systems using single channel processing, we design the multi-channel DSP using Eq.~\ref{eq:Superchannel11} to enable ICI cancellation. The requirements for achieving this can be understood by looking at the limitations of a single channel equalizer, as shown in Fig.~\ref{Fig:Oversamplig}.
The channels are spaced at the Nyquist limit and each channel is shaped with a $\beta=20\%$ raised cosine filter. The part of the spectrum ''seen'' by the equalizer as a function of oversampling or samples per symbol (SPS) is indicated by dashed vertical lines. At 9/8\,SPS the equalizer can almost capture all energy from the channel and at 2\,SPS it has a large spectral margin. The amount of oversampling needed is dependent on the acceptable frequency offset, which causes a non-DC-centered baseband signal. In this work, we choose to design a DSP for 2\,SPS. 

To mitigate ICI, the equalizer need to expand its effective oversampling, similar to how ISI is mitigated by expanding the equalizers temporal memory. 
Looking at Fig.~\ref{Fig:Oversamplig}, this implies that the equalizer has to access more spectral content than what is available to a 2\,SPS equalizer. One way is to use high bandwidth components and sample each channel at a significantly higher oversampling. However, this is not feasible in practise and implies that each channel has to be detected three times. More importantly, high bandwidth analog-to-digital converters (ADCs) typically have worse performance and resolution, preventing this approach in practise~\cite{Laperle2014,Varughese2018}. 

The other option is to exchange information between channels detected individually to effectively realize a high bandwidth receiver. 
This is not possible in traditional systems with independent free-running local oscillator (LO) lasers because uncontrolled relative frequency drift will ''break'' the coherence between channels. 
However, this is possible using frequency combs~\cite{Fontaine2010} and comb-based superchannels can therefore exploit the coherence of the LO comb to effectively create a receiver which spans the full bandwidth of the superchannel, or parts of it, even if each channel is detected using an independent receiver. This technique, known as spectral splizing, is enabled by the intrinsically phase locked lines from a frequency comb~\cite{Fontaine2010}. 
Before going into the details on the joint algorithms, we make an important distinction between different kinds of comb-enabled joint DSP, namely methods only requiring frequency locked carriers vs. methods requiring full phase-locking.
Joint comb-enabled carrier phase estimation (CPE) has been extensively studied~\cite{Yi2010,Millar2016,Lundberg2018,Lundberg2019}. However, joint CPE inherently suffers from dispersive walk-offs up on propagation~\cite{Lorences-Riesgo2016} and requires a fully path-matched transmitter-receiver pair. In addition, using pilot-aided CPE~\cite{Mazur2019d}, the required overhead was only 0.4\%. Reducing this further would therefore have a minimal impact on the overall SE but a high price in terms of distance limitations and path-matching requirements. As such, the algorithms proposed here all only rely on frequency locking rather than phase locking, a very important aspect that also implies that the algorithms are not sensitive to dispersive delay.  

The multi-channel DSP is based on the single channel pilot-based DSP outlined in~\cite{Mazur2019d}. The data is divided into frames consisting of an initial sequence of pilots used for functionalities such as synchronization, FOE and equalization. Residual CPE pilots are then inserted into the payload for continuous tracking. For multi-channel processing, we replace the synchronization, FOE and equalization with a multi-channel equivalent. 
Synchronization consists of a coarse and a fine part. In our multi-channel implementation, this is only done once using a reference channel and this synchronization estimate is then shared among all channels within the superchannel. The synchronization method is equivalent to that for a single channel and described in~\cite{Mazur2019d}. Chromatic dispersion is then compensated individually for each channel. A multi-channel FOE is then used to compensate frequency offsets without breaking the coherence between channels so that stitching is still possible. This step enables the use of independent, free-running, transmitter and receiver combs. Finally, we replace the dynamic equalizer with a multi-channel equalizer to simultaneously compensate for ICI in addition to the standard impairments such as ISI and stochastic polarization effects. 

\subsection{Joint Frequency Offset Estimation}
\label{sec:foe}
The challenge for joint FOE can be understood by expanding the frequency offset $\Omega_i$ from Eq.~\ref{eq:superchannel}. 
For comb-based transmission systems, there are two origins of frequency offsets, which directly corresponds to the two parameters required to fully characterize a comb; the center frequency and the line separation (or repetition rate). The frequency offset $\Omega_i$ in Eq.~\ref{eq:superchannel} can therefore be expanded according to
\begin{equation}
\label{eq:foe}
    \Omega_i = f_0 + i\cdot  \delta f,
\end{equation}
with $f_0$ denoting the difference in center frequency and $\delta f=\Delta f_{\text{Rx}}-\Delta f_{\text{Tx}}$ the difference in line spacing between the receiver and the transmitter comb. 

Importantly, the stitching process relies on knowing the receiver comb spacing $\Delta f_{\text{Rx}}$ 
, not the relative difference $\delta f$ which is the key parameter if one only focus on achieving zero frequency offset. 
This implies that if the frequency offsets are perfectly compensated, stitching cannot be performed unless $\delta f = 0$. 
The joint FOE problem therefore consists of two parts, estimating and compensating for the shared offset $f_0$ and accounting for the  relative difference $\delta f$. 
The magnitude of these offsets depend these depend on the kind of comb used but in general $|f_0| >> |\delta f|$ and estimating $f_0$ is then straight forward using standard methods developed for single channel DSP~\cite{6718054,Mazur2019d}. Multiple channels can also be used for improved accuracy by incorporating knowledge of $\delta f$. 
A tiny spacing difference $\delta f$, a few kHz in this work, is challenging to estimate and requiring long sampling times but also not needed as it can be dealt with directly by the CPE. If $\delta f$ is larger, it is easy to find by comparing estimates from channels spaced further away, and a small correction factor is then applied to each channel to account for the difference. 

The first step is therefore to find and compensate for $f_0$. After this, the received signal or equivalently the input signal to the dynamic equalizer can be written as
\begin{equation}
\label{eq:Superchannel2}
\begin{split}
    r_l(t) = &E_s  \sum_{k=-\infty}^{\infty}s_{l,k} g_{l,l}(t-kT)\exp(j\phi_l) + n(t) + \\ E_s  &\sum_{k=-\infty}^{\infty}s_{l\pm 1,k} g_{l\pm 1,l}(t-kT)\exp(j(\phi_{l\pm 1}  \pm \delta f t)),
    \end{split}
\end{equation}
where we assumed that cross-talk is only present between the closest neighbors. 
If $\delta f$ is large, it will be passed on to add a relative shift for all three channels prior to equalization, or to each channel individually after equalization but before CPE.

\subsection{Multi-Channel Equalization}
The joint dynamic equalization is the key part of this work, focusing on improving performance by mitigating ICI originating from the dense channel spacing. In contrast to previous work~\cite{Pan:12,Liu:13,Liu2012}, our implementation integrates the multi-channel equalization into the standard dynamic equalizer used for polarization tracking and demultiplexing. The equalizer combines information from the channels $l-1,l,l+1$, each one given by the complete Eq.~\ref{eq:Superchannel2}. By accounting for this extra information,  performance can be improved compared to a traditional equalizer. In the frequency domain, this improvement can be understood via Fig.~\ref{Fig:Oversamplig} as giving the equalizer the possibility to use the full information from the side channels to mitigate cross-talk onto the center channel. 
It can therefore coherently superimpose spectral content from both overlap regions, around a normalized frequency of $0.5$ and $1.5$, to distinguish between between the information belonging to the center and to the side channels. However,the effectiveness of this cancellation can be performed naturally depends on the available SNR for the regions under consideration.

\begin{figure}[t]
\centering\includegraphics[width=\linewidth]{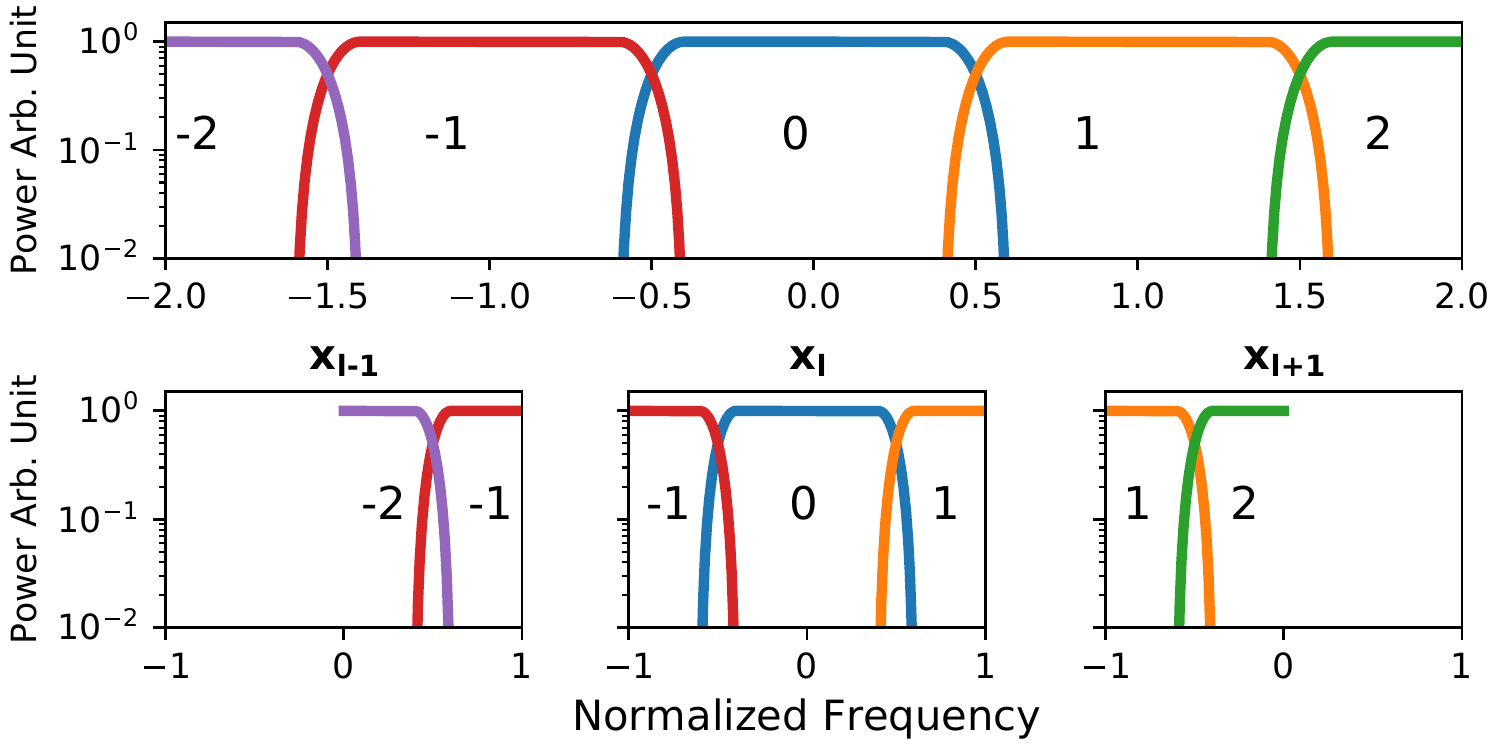}
\caption{\label{fig:aliasing}Illustration of the proposed multi-channel equalizer using aliasing and 2\,SPS. The frequency grid is normalized to the symbol rate. (top) WDM superchannel showing $l\pm2,l\pm1,l$ with channel $l$ to be processed. (bottom) Spectra for input channels, $x_{l-1}, x_{l}, x_{l+1}$  showing the effect of aliasing. }
\end{figure}

This processes is automatically performed by expanding the traditionally used $2\times2$ MIMO equalizer to a $6\times2$ by  including the side channels. The equalization problem can then be expressed as
\begin{equation}
\label{eq:eq_multi_ch}
\begin{bmatrix}
x_{\text{out}}[k]\\ 
y_{\text{out}}[k]
\end{bmatrix} =
\begin{bmatrix} 
\mathbf{w}_{\text{xx},l-1}, \mathbf{w}_{\text{yx},l-1} \\
\mathbf{w}_{\text{xy},l-1},\mathbf{w}_{\text{yy},l-1} \\
\mathbf{w}_{\text{xx},l}, \mathbf{w}_{\text{yx},l} \\
\mathbf{w}_{\text{xy},l}, \mathbf{w}_{\text{yy},l} \\
\mathbf{w}_{\text{xx},l+1}, \mathbf{w}_{\text{yx},l+1} \\
\mathbf{w}_{\text{xy},l+1},\mathbf{w}_{\text{yy},l+1} \\
\end{bmatrix}^T
\begin{bmatrix}
\mathbf{x}_{l-1}\\
\mathbf{y}_{l-1}\\
\mathbf{x}_{l}\\
\mathbf{y}_{l}\\
\mathbf{x}_{l+1}\\
\mathbf{y}_{l+1}\\
\end{bmatrix},
\end{equation}
where $\mathbf{w}$ denotes the different $M$ samples long complex vectors containing the filter coefficients and superscript $T$ the transpose. 
The input channels $l-1,l,l+1$ are sampled at 2\,SPS and we assume that FOE is performed according to the previously outlined method. Finally, as the equalizer in Eq.~\ref{eq:eq_multi_ch} is a linear equalizer, the side channels or their corresponding filter coefficients have to be shifted to reconstruct the frequency grid prior to applying the filter. 

Integrating the coherent stitching into the equalizer has several advantages over first stitching and and then filter the signals using a traditional $2\times2$ MIMO equalizer. First, performing this reconstruction by shifting and directly superimposing the signals would require upsampling of all signals and a core part of this work is to have all processing done at 2\,SPS per channel and upsampling is therefore not allowed. 
If the channels were directly superimposed/stitched without upsampling prior to shifting, this would cause a critical non-revertible performance degradation from aliasing. 
Even neglecting this, direct stitching require knowledge of the exact phase alignment between the different channels and the receiver components need to be fully characterized before detection to avoid penalties from varying frequency response from difference channels~\cite{Fontaine2010}.
The stitching process is made even more cumbersome by the presence of effects such as polarization mode dispersion, varying SNR or small differences in residual dispersion. 
However, by using a dynamic equalizer to perform the stitching, phases and weights are automatically adjusted to achieve the best superposition~\cite{Shi2017}. 

In addition, when using the equalizer structure from Eq.~\ref{eq:eq_multi_ch}, aliasing can be turned into something advantageous rather than a drawback. 
This is because the side channels are separate inputs and the superposition is performed by the equalizer, which cannot tell the difference. This implementation avoids the need for resampling before equalization. The input superchannel is shown in Fig.~\ref{fig:aliasing}(a) with the resulting equalizer inputs for the lower, center and upper input channels shown in Fig.~\ref{fig:aliasing}(b)-(d). For the upper and lower bands in Fig.~\ref{fig:aliasing}(b) and (d), we observe that the side channels appear on the intuitive ''wrong'' side compared to Fig.~\ref{fig:aliasing}(a), showing a clearly aliased signal. 

\section{Experimental Setup}
\label{Sec:Setup}
\begin{figure}[t]
\centering\includegraphics[width=0.5\textwidth]{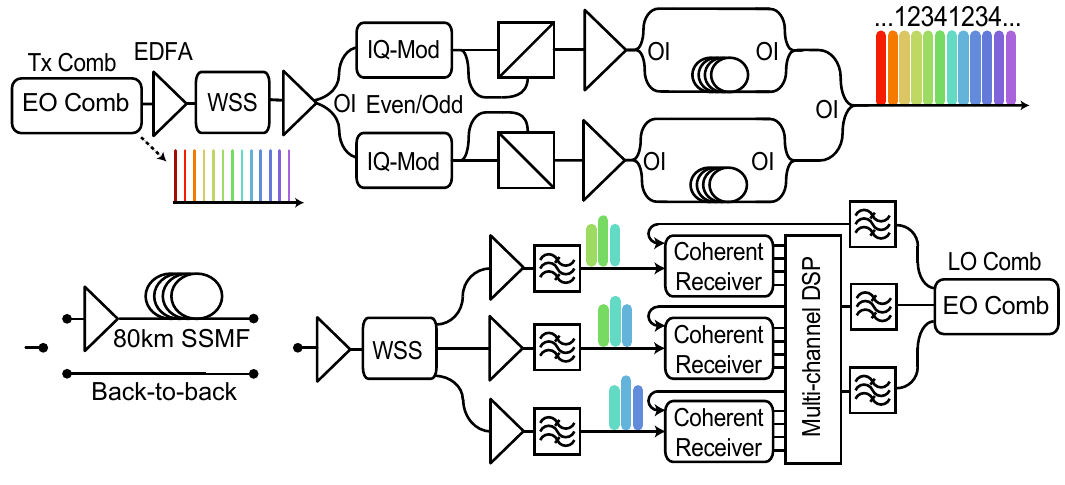}
\caption{\label{Fig:Setup} Schematic of the experimental setup used for multi-channel detection and joint signal processing. EO Comb: Electro-optic frequency comb, EDFA: erbium-doped fiber amplifier, WSS: wavelength selective switch, OI: optical interleaver, AOM: acousto-optic modulator, SSMF: standard single-mode fiber, LO: local oscillator. }
\end{figure}

The experimental setup used for system-level evaluation of the proposed multi-channel DSP is shown in Fig.~\ref{Fig:Setup}. We used two separate electro-optic (EO) combs as our transmitter and LO comb, respectively. Each comb was fully independent and built with a phase modulator and an intensity modulator, generating about 21 lines with 5\,dB flatness. The modulators were driven using two independent, 25\,GHz RF-clocks and seeded using two external cavity lasers with about 100\,kHz specified linewidth. Frequency noise measurements~\cite{Kikuchi2012} resulted in an equivalent Lorentizan linewidth of about 10\,kHz with significant presence of both 1/f-noise and Gaussian noise~\cite{Mercer1991}. 

A 5.5\,dB noise figure EDFA amplified the transmitter comb output before a wavelength selective switch (WSS) was flattened its output. Cascaded 25\,GHz optical interleavers (OIs) then separated the lines into even/odd. These were modulated independently with two IQ-modulators driven by two digital to analog converters (DACs) each. Polarization multiplexing (PM) signals were emulated using the split-delay-combine method with about 250\,symbols delay for each modulator. Prior to recombining, an additional set of OIs was used to introduce further decorrelation between every second even/odd channel, respectively, producing a 1-2-3-4-1-2 decorrelation scheme. The signals were then recombined and amplified to form the final superchannel. 
\begin{figure}[t]
\includegraphics[width=0.5\textwidth]{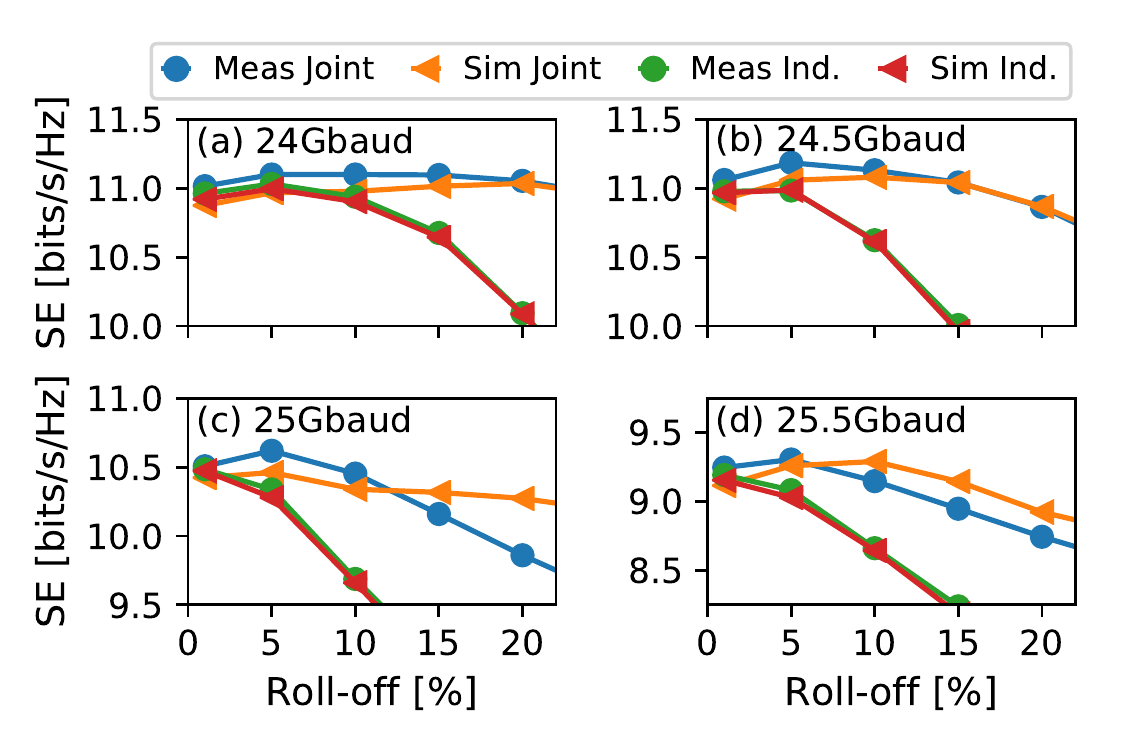}
\caption{\label{Fig:B2B_1}Comparison of single channel and joint processing in back-to-back configuration for 64QAM. The measurement numerical simulations, showing a good agreement with the measurement results. AIR as a function of $\beta$ for 24,\,24.5,\,25 and 25.5\,GBaud is shown in (a)-(d), respectively. \vspace{-3mm}}
\end{figure}
For B2B characterization, the transmitter was directly connected to the receiver.
Transmission was demonstrated with a span of  80\,km standard SMF and about 16\,dB spann loss followed by an EDFA with  $\sim5.5$\,dB noise figure. The amplified output was then connected to the receiver. 

The receiver consisted of a pre-amplifier followed by a multi-port WSS to select three adjacent channels from the superchannel with the number of detected channels being limited by equipment constraints. The filtering bandwidth for each channel was set to 0.3\,nm. The outputs were independently amplified and fed to three separate coherent receiver. Three corresponding LO-lines from the receiver comb were selected using a cascade of OIs and filters. The electrical outputs were sampled using three 4-channel real-time oscilloscopes operating at 50\,GS/s each. Offline DSP was implemented using the joint functionality presented in Section~\ref{sec:joint_dsp}. Each channel was also processed individually using the DSP presented in~\cite{Mazur2019d}. Performance was measured by estimating the generalized mutual information (GMI)~\cite{Alvarado2016} using about $10^6$\,bits/batch and averaging over 4 batches. The SE was calculated by dividing the GMI with the guardband overhead. The DSP pilot overhead was optimized to maximize the AIR, resulting in an optimal sequence length of 2048 symbols and an optimal CPE insertion ratio of 1/256, corresponding to 0.4\% overhead. The frame length was limited by the available DAC memory to 108800 symbols, resulting in a total overhead of 2.3\%.  

\section{Results}
We investigated the effect of cross-talk mitigation using our comb-enabled joint DSP by varying the symbol-rate and pulse-shaping roll-off in a back-to-back (B2B) experiment. The performance and potential of our technique was further varified in a transmission experiment.

\subsection{Back-to-Back Characterization}
The measured GMI as a function of roll-off factor for 24,\,24.5,\,25 and 25.5\,GBaud symbol-rate on a 25\,GHz grid is shown in Fig.~\ref{Fig:B2B_1}(a)-(d). Here we take average of the three channels for the single channel performance (relative performance differed with about 0.1\,bits/4D-symbol). The measured results are compared to Monte-Carlo simulations. Each point is an average of 5 simulation runs, consisting of $2^{18}$ 4D symbols each. To match the experimental condition, the simulation model accounted for limited effective number of bits (ENOBs) (4.5\,bits) and background ASE from amplifiers in the transmitter was modelled by adding AWGN to an SNR of 35\,dB, resulting in a good match between measured and simulated single channel performance. The SNR value also match well with the estimated OSNR of the transmitter measured to be 37\,dB at 25\,GBaud. 
A random frequency offset $0 \leq |f_0| \leq 2$\,GHz and an independent polarization rotation for each channel were added but did not affect the performance as it was directly removed by the pilot DSP. Extensive numerical simulations were also used to verify that the decorrelation system was sufficient by comparing the scenarios of fully independent channels to the decorrelation scheme, showing no difference in performance. 

\begin{figure}[t]
\includegraphics[width=0.5\textwidth]{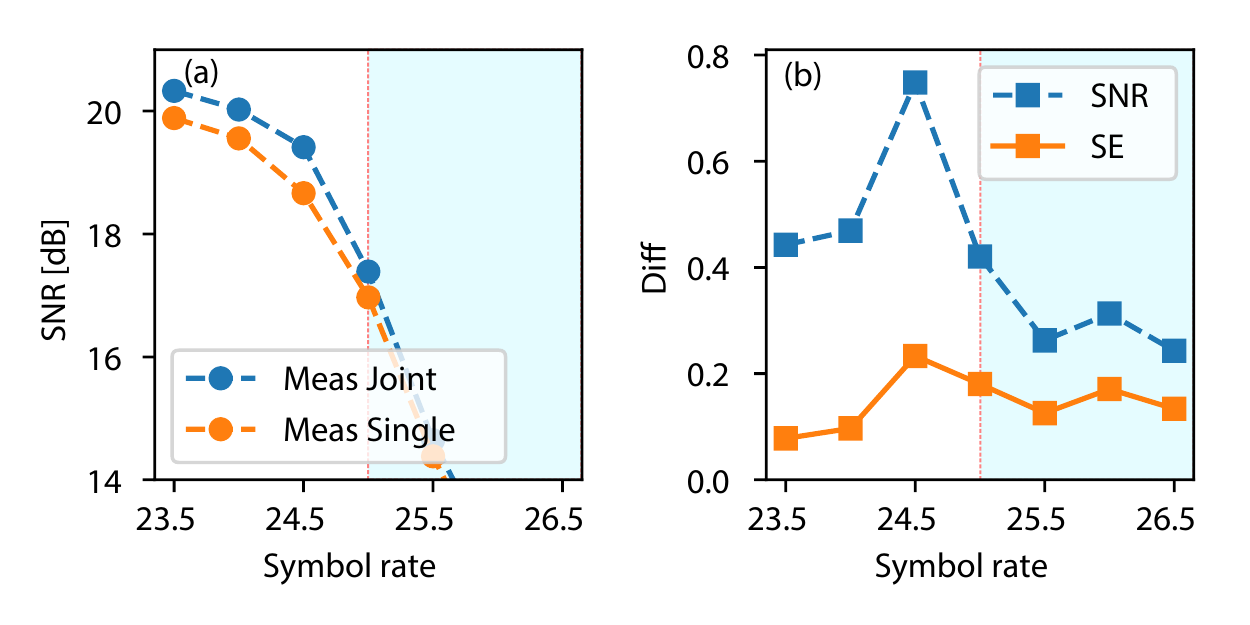}
\caption{\label{Fig:B2B_2}(a) SNR comparison between joint and individual processing for different symbol rates. (b) Resulting difference in SNR and SE. Regions corresponding to super-Nyquist spacing are indicated in blue.  }
\end{figure}

Figure~\ref{Fig:B2B_1} shows several interesting, and from a theoretical point of view, unexpected results. Firstly, we observe that the joint-DSP measurements outperform the simulations at low roll-off factors. Secondly, we observe that a small amount of crosstalk is beneficial for the system, as the performance with a roll-off $\beta=5\%$ is higher than that with $\beta=1\%$, which in theory should have no crosstalk for symbol rates of 24 and 24.5\,GBaud. This is because the DACs and ADCs are not ideal, and as $\beta \rightarrow 0$, the peak-to-average power ratio (PAPR) increases~\cite{Daumont2008}, causing a significant penalty due to the limited ENOBs~\cite{Varughese2018}. For single channel processing, this is dominant as linear crosstalk (even if it is just a few \%) will degrade the signal quality rapidly. However, for joint EQ, this cross-talk can be mitigated and new optima arise. For 24.5\,GBaud, as shown in Fig.~\ref{Fig:B2B_1}(b), we observe that the optimal point is $\beta=5\%$ but even at $\beta=10\%$, the performance is better than at $\beta=1\%$ for the single channel case. The this trend is maintained as the symbol rate is increased and even at 25.5\,GBaud (SuperNyquist). 

These results are further verified in Fig.~\ref{Fig:B2B_2}. The measured SNR is shown as a function of channel symbol rate in Fig.~\ref{Fig:B2B_2}(a), displaying a noticeable difference between the two curves. For each point, the optimal $\beta$ from Fig.~\ref{Fig:B2B_1} has been selected and we emphasize that for the joint processing, the optimal $\beta$ is always larger than 1\%. 
Looking at the difference between joint and individual processing, shown in Fig.~\ref{Fig:B2B_2}(b), we observe that the SNR gain from joint equalization peaks at about 0.75\,dB for 24.5\,GBaud, a very large gain given the closeness to theoretical bounds of the investigated system. At this point, the SE increase is 0.25\,bits/4D symbol. For the complete 21 channel superchannel, this corresponds to an increase in total throughput of 130\,Gbit/s and the relative SNR improvement is 5\%. Effectively, the equalizer shifts the optimal symbol rate from 24 to 24.5\,GBaud in a penalty-free way and the gain in throughput matches this shift. 

\begin{figure}[t]
\includegraphics[width=0.5\textwidth]{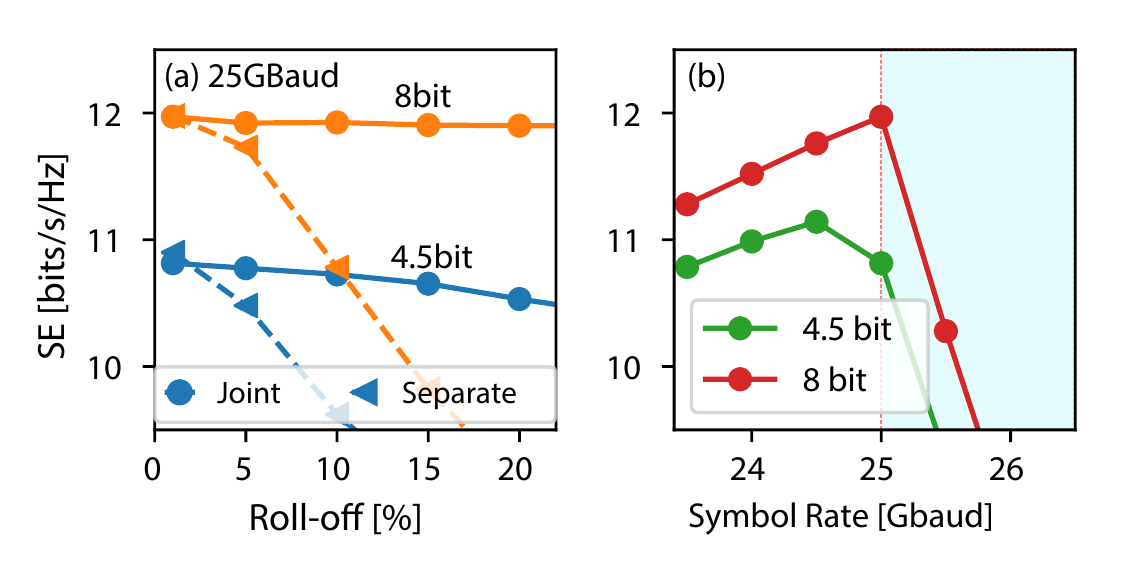}
\caption{\label{Fig:B2B_3}Comparison between a transceiver-limited system having 4.5\,ENOBs and a system with an 8\,ENOBs for 64QAM. (a) Comparison between joint and individual processing for the two cases at 25\,GBaud. The shape is very similar between the two cases but an absolute offset is found. (b) SE as a function of symbol rate for the two cases showing that combining the joint processing with resolution, the optimal position changes to zero guardband.   }
\end{figure}

The ideal goal is of course to use all the available spectrum to transmit information and therefore have no guard-bands within the superchannel. The reason for the residual guard-band is that the effective SNR is not high enough, so that noise and especially quantization noise degrades the performance of the system, and the relative gain in SNR from the joint equalizer. This is shown in Fig.~\ref{Fig:B2B_3}, comparing the previous simulation results and simulations with 8\,ENOBs to the estimated performance of a system using higher resolution DACs and negligible noise. Figure~\ref{Fig:B2B_3}(a) shows a comparison at zero-guardband (25\,GBaud) and Fig.~\ref{Fig:B2B_3}(b) shows the achieved SE versus symbol rate with the optimal roll-off selected for each point. From Fig.~\ref{Fig:B2B_3}(a), we observe that there is a gap in terms of absolute performance and that the shape and difference between joint and individual processing is very similar in both cases. The increase in resolution can therefore be exploited to close the gap by allowing for removing the guard-bands, as shown in Fig.~\ref{Fig:B2B_3}(b). 

\subsection{80 km Transmission}
\begin{figure}[t]
\includegraphics[width=0.5\textwidth]{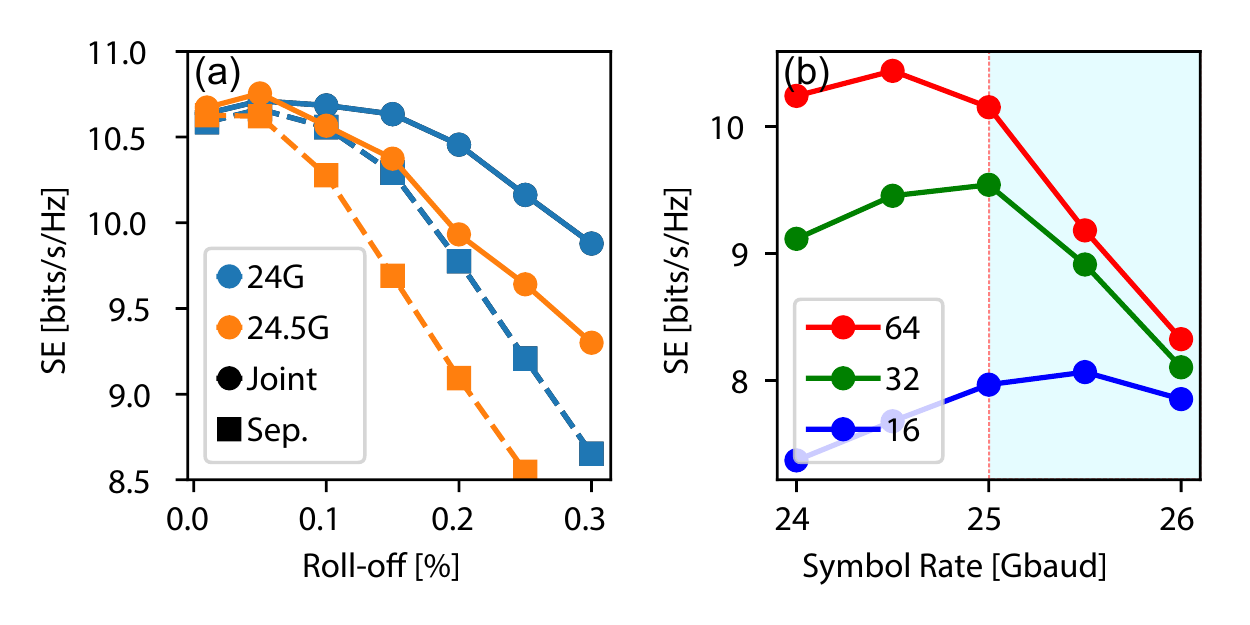}
\caption{\label{Fig:80km_1}(a) SE for joint and individual equalization at 24 and 24.5\,GBaud after transmission through an 80\,km SMF span at optimal launch power of 9\,dBm. (b) AIR as a function of symbol rate for 16,\,32 and 64QAM, super-Nyquist spacing is indicated in blue. }
\end{figure}

Finally, we verified that the proposed system is compatible with the stringent requirements of fiber transmission. Experiments where performed at the optimum launch power of 9\,dBm, corresponding to about -5dBm/ch. First, we compared the performance of the joint equalization with single channel while varying $\beta$ for 24, and 24.5\,GBaud as shown in Fig.~\ref{Fig:80km_1}(a). However, as previously noted, this depends largely on the ratio between the different noise components and the requirements for the selected modulation format. Figure~\ref{Fig:80km_1}(b) shows the SNR for different symbol rates at optimal $\beta$ using the joint processing for 16,32 and 64QAM. Here, the effect of optimal guard-band is clearly visible as we observe that 16QAM has its optimal AIR at 25.5\,GBaud, corresponding to a faster-than-Nyquist system~\cite{Mazo1975}. These systems traditionally use minimum distance sequence detection to combat the severe ISI/ICI~\cite{Colavolpe2011}, and in this case, the proposed joint equalizer can improve the performance. 

\begin{figure}[t]
\includegraphics[width=0.5\textwidth]{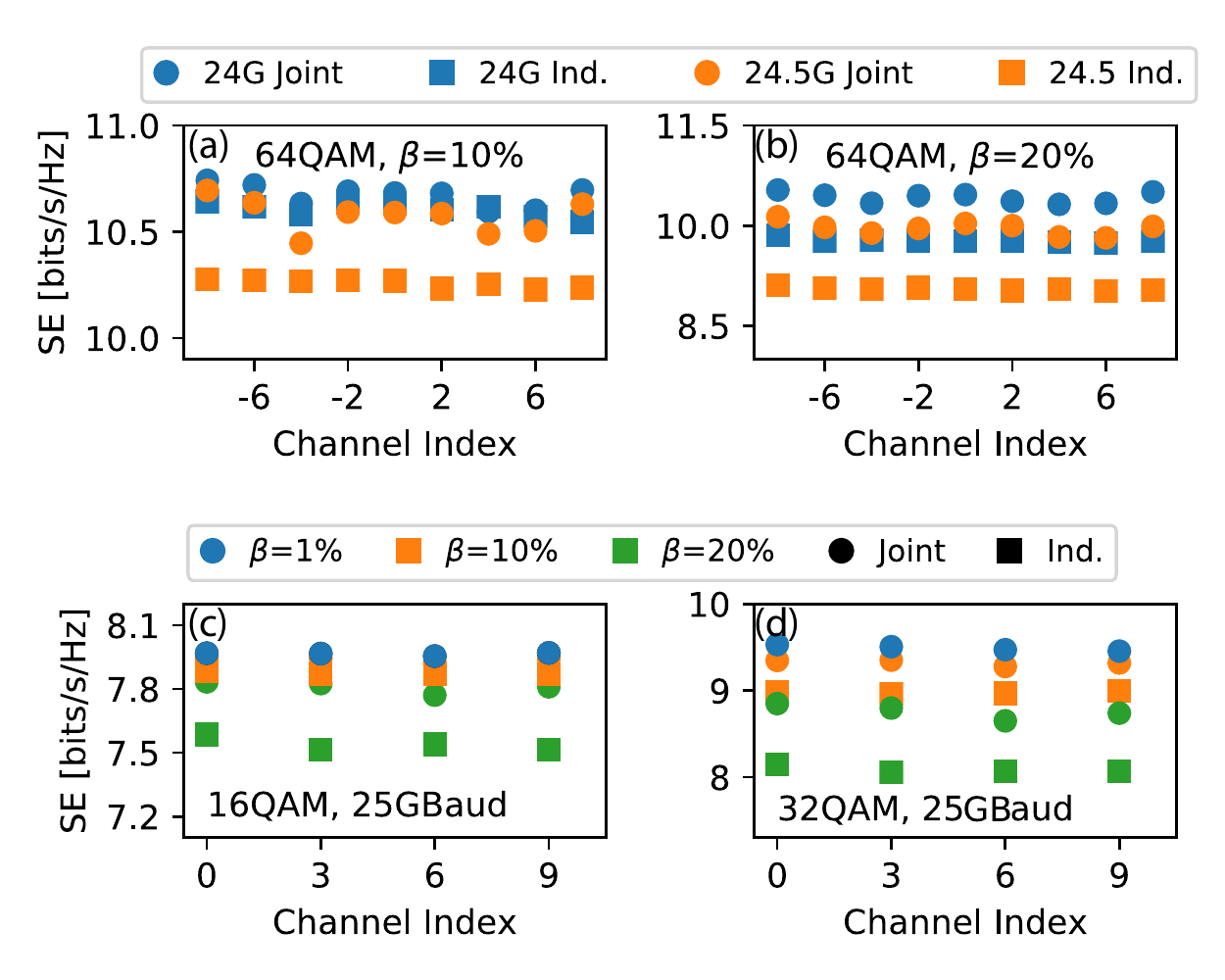}
\caption{\label{Fig:80km_2}Measured SE for 64QAM with (a) $\beta=10\%$ and (b) $\beta=20\%$ for selected test channels evenly spaced within the superchannel at symbol rates of 24 and 24.5\,GBaud. Measurements at 25\,GBaud for varying roll-off factor using 16QAM and 32QAM are shown in (c) and (d), respectively. }
\end{figure}

Finally, we measured the performance of multiple evenly-spaced channels within the superchannel to verify that successful processing is possible after transmission, independent of the channels spectral position. 
the performance of all channels to verify performance beyond the center of the combs. 
The results for 64QAM with 24 and 24.5\,GBaud are shown in Fig.~\ref{Fig:80km_2}(a) for $\beta=10\%$ and (b) for $\beta=20\%$. Every second channel was measured, neglecting the edge channels as they only have one neighboring channel. We observe that, although the performance within the superchannel in absolute SE, the gain from joint processing is fairly constant. In Fig.~\ref{Fig:80km_2}(a) for 24.5\,GBaud, we observe a large difference between joint and individual equalization. When additional cross-talk is induced by increasing the roll-off to 20\%, as shown in Fig.~\ref{Fig:80km_2}(b), we observe a large and clear difference between joint and individual processing, being 0.7 and 0.9\,bits for 24 and 24.5\,GBaud, respectively. The performance for 16 and 32QAM at 25\,GBaud is shown in Fig.~\ref{Fig:80km_2}(c) and (d), respectively. 
Here we observe that the difference between a $\beta=1$ and 10\% was below 0.1\,bit in SE when using joint processing, but the cross-talk penalty rapidly reduces the SE when each channel is processed individually. Combining the results from Fig.~\ref{Fig:80km_2}, we observe that the performance for joint processing with $\beta=20\%$ is very close to that of individual processing using $\beta=10\%$, despite the fact that the overlap bandwidth is doubled. A similar trend is observed between $\beta=1$ and 10\% but the relative difference is smaller. 

\section{Discussion}
The results presented clearly show that by realizing superchannels using frequency combs, unique comb properties can be exploited to realize performance which exceeds that of traditional superchannels. 
Only looking at the spacing, DSP-based locking of the individual lasers is an alternative, realizing $<$100\,MHz line stability~\cite{Rahn2018}. While this is orders of magnitude larger than the relative drift of high-end frequency combs, it is enough to drastically reduce the overhead from inter-channel guardbands. Although sub-MHz stability is not needed for this, it is needed to allow for spectral stitching and the architecture presented in this work. This intrinsic line stability, defining a frequency comb, is therefore a necessity  to reach the ultimate efficiency using practically realizable components. 

The core issue with uncompensated crosstalk can also be seen by comparing the recent SE records using formats reaching 4096QAM~\cite{Terayama2018,Olsson2018a}. 
While the record potential single channel SE is >19.77\,bits/s/Hz~\cite{Olsson2018}, despite using a highly stable EO comb and equivalent hardware, the SE was reduced to 17.3\,bits/s/Hz for a superchannel~\cite{Olsson2018a}. The superchannel consisted of 10$\times$3\,GBaud channels with optimized guard-band overhead of 5\%. The small symbol rate is necessary to cope with the extreme resolution and ENOB requirements for PS-4096QAM but the efficiency is still drastically reduced by the guard-bands. Expanding such a system with the proposed multi-channel DSP, the guard-bands can be further reduced and much higher ENOBs and SNR will improve the ICI mitigation performance compared to the results presented here. 

Another key aspect is how to achieve the required RF clock synchronization between the transmitter and receiver comb without requiring DSP-based estimation of $\delta f$, which is increasingly cumbersome when $\delta f \rightarrow 0$ as discussed in Section~\ref{sec:foe}. 
The RF clock driving the receiver comb can be synchronized to the sampling clock running the ADCs on all channels. This is very important as it makes stitching process transparent to small drifts in $\Delta f_{\text{Rx}}$. If a different kind comb source such as a microcomb with spacing defined by a cavity~\cite{Herr2013,Suh2018} is being used, the spacing can be extracted from the beat note of two or more lines. A phase-locked loop can then be used to filter the signal~\cite{Mazur2018b}, which is then multiplied and used as the ADC sampling clock. This implies that spectral slizing can be done without any DSP-based estimation of $\delta f$. 
Depending on stability, these RF-clocks could either be free-running (the commercial RF-clocks used in this work had a relative stability of $|\Delta f_{\text{Rx}}-\Delta f_{\text{Tx}}| < 5$\,kHz over days). Otherwise, preferably to ensure long term stability, the RF clocks could be references externally using for example cheap, off the shelf, components for GPS locking with sub-Hz accuracy~\cite{Lombardi2008}. 

A key question is of course the added complexity of multi-channel equalization. 
Hardware implementations are far beyond the scope of this work aiming at demonstrating the principle. However, designing the multi-channel DSP to not require upsampling is an example of a key feature that will ease a hardware implementation.
Similarly, the dispersion compensation filters are jointly designed to ensure a smooth filter response also in the overlap regions but each channel is filtered independently. Even for single-channel processing, dispersion compensation accounts for about one third of the total DSP power consumption~\cite{Pillai2014} and as joint filtering is not needed, single channel processing is preferred to reduce complexity. 
However, joint processing also bring benefits in terms of reducing complexity of other modules. One example is the required pulse-shaping filter length, which can be drastically reduced when the optimal roll-off increases. 

Finally, we note that this work only scratches the surface of the potential for resource sharing and multi-channel processing for comb-based superchannels. One example, which is directly compatible with the proposed linear DSP, digital back propagation to mitigate non-linear distortions~\cite{Liga2014}. This can be performed at the transmitter side by exploiting the frequency locking of the transmitter comb~\cite{Temprana2015,Temprana2015a} and at the receiver side by means of coherent stitching~\cite{Fontaine2013a}. Hybrid approaches are also possible but have, to the best of our knowledge, not yet been investigated. 
Similarly, if the RF clocks are not very precise between the transmitter and receiver, DSP-based clock recovery is needed and standard methods cannot cope with very aggressive roll-off factors~\cite{Wang2003a}. This issue is not limited to clock recovery and also found for dispersion estimation~\cite{Yao2015}, to mention one example. By exploiting the proposed joint processing to allow for increasing the roll-off without degrading the overall performance, accurate DSP-based estimation is enable. This highlights that joint DSP has a large potential to enable new trade-offs in the design of superchannel systems. 

From a more communication-theoretic perspective, a superchannel also provides a structured way or realizing a high-dimensional signal space. Assuming that joint processing and information exchange between channels is performed in the transceivers, modulation and coding can be expanded to exploit this new signal space~\cite{Millar2014}. This can be used to increase the sensitivity of modulation formats and FEC can be redistributed and shared among the channels. This could be exploited to allow for very long codes to be used without suffering from increased waiting time, similar to coding over multiple spatial dimensions~\cite{Puttnam2015}, or to allow for interleaving within the superchannel to reduce its time memory, to mention two examples. 

\section{Conclusions}
We have presented a comb-based superchannel exploiting the intrinsic line stability of optical frequency combs to enable multi-channel processing. 
In addition to overcoming the relative frequency drift from free-running lasers, the locked frequency grid is exploited to extend the dynamic equalizer to also mitigate inter-channel interference and thereby shift the optimal guardband towards zero. All channels are received individually and sampled at two samples per symbol. Rather than upsampling, the coherent combining is directly performed by the dynamic equalizer by exploiting aliasing to overcome the upsampling requirement. Experimental results in back-to-back configuration and after 80\,km transmission of a 21 channel superchannel originating from a 25\,GHz-spaced optical comb show that the effective cross-talk mitigation, reducing the optimal guard-band from 1\,GHz to 500\,MHz. It furthermore changes the optimal spectral shape from 1\% roll-off towards 10\%, making systems using multi-channel processing inherently more tolerant to transceiver limitations. In addition, the larger roll-off reduces the required equalizer temporal memory and simplifies DSP-based parameter estimation. The effective improvement in received signal to noise ratio enables higher throughput or extended reach, showing that comb-based superchannels can realize increased system efficiency in ways that are directly relying on the intrinsic line stability of the comb and therefore not possible using traditional laser arrays.

\section*{Acknowledgment}
The authors would like to thank Dr. Benjamin Foo for fruitful discussions. Funding from the Swedish scientific council (VR) via grants 2015-04239 and 2017-05157 and the Knut and Alice Wallenberg foundation.

\ifCLASSOPTIONcaptionsoff
  \newpage
\fi

\bibliographystyle{IEEEtran}
\bibliography{Mazur_References_Cleaned}

\end{document}